\documentclass[conference]{IEEEtran} 

\usepackage{graphicx,subfigure}
\usepackage{color,listings,url}

\begin{document}

\title{Irida: A real-time Wireless Sensor Network visualization feedback protocol\footnotemark[1]}

\author{
Marios Karagiannis~\footnotemark[2], Laetitia Dallinge~\footnotemark[2] and Jos{\'e} Rolim~\footnotemark[2] \\
\small \footnotemark[2] {\rm Centre Universitaire d' Informatique, Geneva, Switzerland.}\\
\small E-mail: 
{\tt \{marios.karagiannis, jose.rolim\}@unige.ch,\tt dalling0@etu.unige.ch}

\\

}
\maketitle

\begin{abstract}
  
\\
\noindent
\\
In this paper, we describe the implementation of a real time visualization and feedback system for Wireless Sensor Network algorithms. The system is based on a fixed hardware testbed, which is deployed on a vertical flat surface and a feedback loop system that takes information about the current state of the network and projects this state, in a visual way, on the surface itself using a video projector. The protocol used is open and simple to use, and can be easily adapted for different hardware configurations. We call our system Irida.

\end{abstract}

\section{Introduction and Related Work}
\label{sec:intro}
Wireless Sensor Networks (WSN) are autonomous networks of small devices that feature environment sensors and wireless communication capabilities. The main characteristics of these devices(also called sensor nodes or motes) are their limited computation capabilities and limited power capacities. Networks formed using sensor nodes do not generally require external infrastructure and are self organizing. WSN algorithm research is currently a research hot topic because of the uniqueness of the requirements of such algorithms, which are in most cases ad-hoc and focus on energy saving. Traditionally, information on WSNs are captured from the network nodes' sensors, stored and processed locally and then transmitted using the built-in RF capabilities to central stations, called sinks.

\par

Due to the nature of wireless sensor networks, namely the large number of nodes in the network and the wireless factor, it has always been difficult to monitor algorithm functionality and assess their performance for Wireless Sensor Networks while using real hardware. For this reason, simulation has been a popular choice among researchers for WSN algorithms. Although simulating the behaviour of an algorithm can have advantages, such as re-playability and fine tuning of conditions, ultimately the algorithms have to be evaluated in a real environment since this is where they are destined for. In most cases, researchers, along with the development of the network or other algorithms for the network, they have to develop ad hoc tools for visualizing the execution of such algorithms, especially in cases where no actuators are part of the set-up. Although, for final installations this might not be necessary in all cases, it is definitely useful for development, testing, fine tuning, demoing and teaching purposes in almost all cases. In this context, we believe that although simulation is a good first step for developing WSN algorithms, it should not be the last one as well.

\par

Our contribution is a feedback loop visualization system, which we call Irida. Irida attempts to provide a set of tools to help ease the evaluation on real hardware and also to demo the results of wireless sensor network algorithms. Irida consists of 3 parts:
\begin{itemize}
\item The Irida enabled sensor firmware. This is the hardware independent implementation of the Irida protocol. It should work along with the actual algorithm tested or demoed and provide the necessary feedback to the next module.
\item The Irida Control Unit (ICU). This module is responsible for gathering all the information from the previous module and forwarding them to registered visualization modules.
\item The Irida Visualization Unit (IVU). This module registers with an ICU, receives feedback data and visualizes the network. IVUs can take many forms, depending on requirements (projection units, mobile devices, desktop computers etc.). More than on IVU can be attached to a given Irida system, providing real time feedback to local and remote locations alike.
\end{itemize}
The system is destined to be a valuable tool for researchers in the area of Wireless Sensor Networks. In order to utilize the system to evaluate and/or demo WSN algorithms, the first step that has to be taken is to insert the Irida protocol side by side with the actual algorithms while developing the nodes firmware. This is independent of the make of the hardware used, as long as it provides a way to forward the Irida packets to the ICU associated to the WSN. This can be through USB, serial or even wirelessly alongside with the other wireless packets the networks exchange for the needs of the algorithm or algorithms being tested. As long as the nodes can forward visualization commands to the corresponding ICU, the system will be able to visualize the network status and activity using Irida.
\par
We should note that the protocol used to communicate between the network and the ICU is identical to the protocol used to communicate between the ICU and the IVU or IVUs. The only difference is the connection type (ZigBee, USB, Serial for the WSN to ICU connection versus UDP between the ICU and IVU or IVUs). Because of this fact, in this stage, where the system is used solely for visualization purposes, the system can actually work even with the omission of an ICU, with the WSN nodes talking directly to an IVU. We are proposing the introduction of an ICU as a mediator and as a constant element of the system, in order to provide modularity and flexibility, to make it possible for IVUs to register to the ICU on demand without limitations on the type or number of IVUs that can use the system at the same time. On top of that, we plan to expand the functionality of the ICU in the future, as explained in section~\ref{sec:future}.

\par

Because of the Irida architecture choices, there is no real limitation on the source of this data, and it could very well be from simulated networks as well, but most simulated environments already have their own ad-hoc visualization solutions(for example WSNGE\cite{karagiannis09} or Shawn\cite{kroeller05shawn}). For specific hardware platforms, there are some solutions also (for example Surge Network Viewer\cite{surge} or MoteView\cite{Turon:2005:MSN:1251990.1253395} for Crossbow's\cite{crossbow} motes or TinyViz\cite{Levis:2003:TAS:958491.958506} for TinyOS\cite{tinyos} and TOSSIM\cite{Levis:2003:TAS:958491.958506}). Irida is designed to be more of a solution for a generic real hardware visualization protocol. 

\par
SpyGlass\cite{Buschmann:2004:STC:1031495.1031546} is another visualizer built with flexibility in mind. SpyGlass uses plugins for drawing primitives, record and playback network activity. SpyGlass is using a similar architecture to the one we propose for Irida, although we extend some of the ideas of SpyGlass to allow even more flexible configurations and we also focus more on the visualization data feedback protocol, than on the visualization itself with our work. SpyGlass is a visualization software package. Irida is a generic visualization protocol that consists of independent modules, where the visualization units (IVUs) are interchangeable and just a part of the system. In fact, SpyGlass, or any other visualizer that accepts on-line data for displaying, could be modified to utilize Irida as a visualization protocol option, thus turning into a fully functional IVU with minimal effort.

This work is partly implemented to be used during the life time of the Hobnet project. The Hobnet project (June 2010-May 2013)\cite{hobnet} is a project that uses existing FIRE technologies and platforms for Future Internet applications focused on automation and energy efficienty for smart/green buildings. The main research areas of Hobnet are the use of IPv6/6LoWPAN infrastructure, 6lowApp standarization, new algorithmic models that fit the objectives of the project and are scalable and energy efficient, rapid development and integration for building management and support for deployment and monitoring of the applications on existing FIRE testbeds. Irida will be used to evaluate and demo some of the algorithms and scenarios that will be developed in the lifetime of this project.

\begin{center}
\begin{figure}
\includegraphics[scale=0.35]{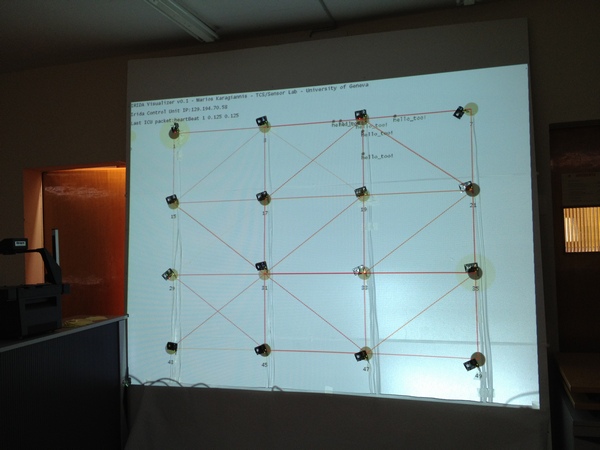}
\caption{4x4 grid testbed running Irida (controlled neighbourhood discovery)}
\label{fig:irida4x4_1}
\end{figure}
\end{center}
\begin{center}
\begin{figure}
\includegraphics[scale=0.35]{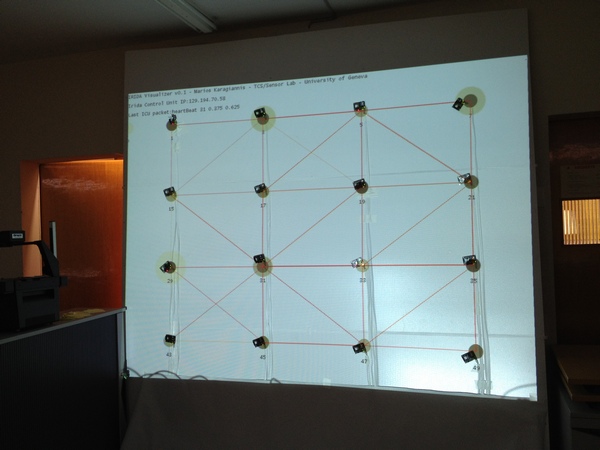}
\caption{4x4 grid testbed running Irida (controlled neighbourhood discovery)}
\label{fig:irida4x4_2}
\end{figure}
\end{center}

\begin{center}
\begin{figure}
\includegraphics[scale=0.35]{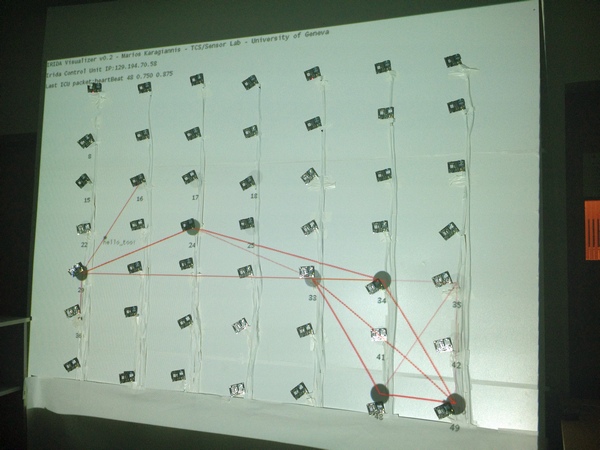}
\caption{7x7 grid testbed running Irida (free neighbourhood discovery) with sleeping nodes}
\label{fig:irida7x7_1}
\end{figure}
\end{center}
\begin{center}
\begin{figure}
\includegraphics[scale=0.35]{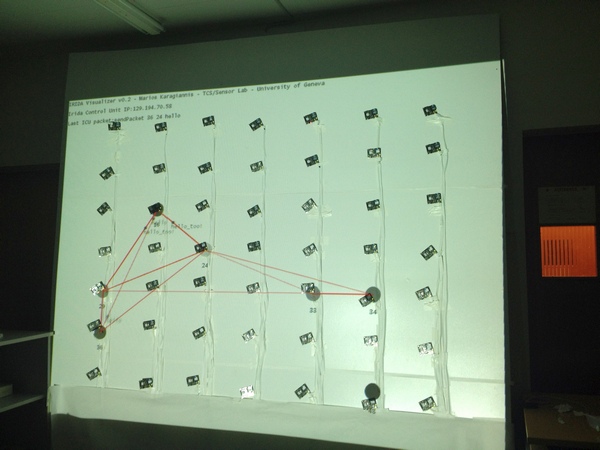}
\caption{7x7 grid testbed running Irida (free neighbourhood discovery) with sleeping nodes}
\label{fig:irida7x7_2}
\end{figure}
\end{center}
\section{Hardware used}
\subsection{Testbed}
The implementation on our lab was made possible by hardware from Libelium\cite{libelium}. To test the applicability and scalability of out system, we created a grid based testbed consisting of 16 sensors in a 4x4 grid (fig.~\ref{fig:irida4x4_1} and fig.~\ref{fig:irida4x4_2}) and then a grid of 49 sensors forming a 7x7 grid(fig.~\ref{fig:irida7x7_1} and fig.~\ref{fig:irida7x7_2}). For both configurations we used Waspmote sensors from Libelium\cite{libelium}. The Waspmote sensors use an ATmega1281 micro-controller at 8MHz, 8KB or SRAM, 4KB of EEPROM and 128KB of Flash memory. They also can utilize SD Cards of 2GB. They weight only 20gr which came in handy, because of the nature of the deployment. Waspmotes have the ability to plug in one or two modules with Zigbee, Zigbee PRO, Bluetooth or GPRS radios. For our testbed, we used Zigbee\cite{zigbee} with a built-in antenna (txPower of 1mW and sensitivity -92dB), in order to be able to change the radio power using the firmware (the theoretical maximum range for this configuration is 500m which is too much for our purposes). This was also optimal, since our deployment was done in one room and larger antennas were non needed. All nodes were powered through the built-in USB interface. The USB interface was also used to route debugging message that controlled the visualization.
\par
Libelium provides USB interfaces for PC-to-network communication with matching communication modules. Another important aspect that drove us to choose the Waspmote platform, is the fact that Waspmotes can receive new firmwares wirelessly, store them in a built-in SD card and execute them on demand. More than one firmwares can be stored at any given time and wireless distribution of new firmwares can be done using unicast, multicast or multihop OTA mechanisms.
\section{Implementation aspects}
\subsection{The reporting pipeline}
\begin{center}
\begin{figure}
\includegraphics[scale=0.4]{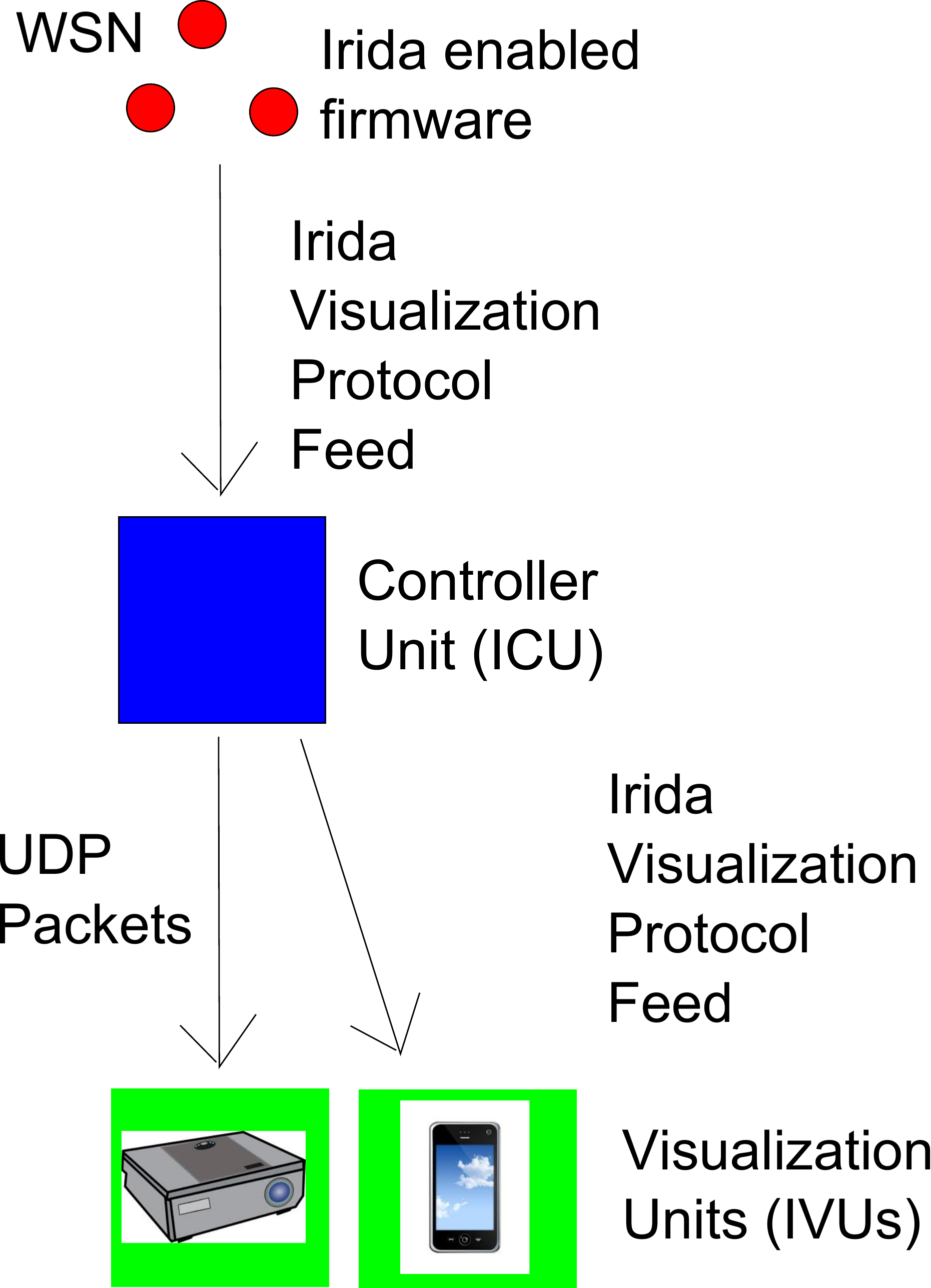}
\caption{Irida architecture}
\label{fig:irida_feedback_flow}
\end{figure}
\end{center}

We have designed Irida to be a multi-module system. As long as each module follows the Irida protocol, modules could easily be expanded and are interchangeable. The basic idea for the reporting pipeline is shown on figure~\ref{fig:irida_feedback_flow}. 
\par
Information starts on the nodes of the WSN firmware. The sensors use the protocol described in section~\ref{sec:protocol} to send the information they need visualized to the Irida Controller Unit (ICU). This can be done in a wireless fashion (using the same wireless module they use for their operation or a secondary one) or using a wired connection (for example a direct USB or serial connection between the sensors and the computer running the ICU). The ICU maintains a list of IVU that have previously registered with it. Each IVU can connect and disconnect to the ICU at any time and start receiving visual information. This provides great flexibility to the Irida system, since IVUs can be as static as a PC with a wall projector connected using the LAN or as dynamic and mobile as a smart phone or tablet connected through the Internet to the ICU. In both cases, the IVU will have to register with the ICU and, provided it has implemented a display engine that responds to the protocol in section~\ref{sec:protocol}, it will provide visual feedback in real time.

\subsection{The reporting protocol}
Visualization commands are given directly from the node firmware and are heavily application dependant. The visualization commands are gathered by the controller and forwarded to registered visualizers through UDP connections. The system was primarily designed so that the controller and visualizer are at close proximity, so UDP was chosen for simplicity reasons. UDP is connectionless and does not guarantee packet order or packet delivery. Packet order is not important in this scenario and connection maintenance is not needed. Even if the protocol is used of a more unreliable connection (for example over GSM), most protocol instructions are redundant, so that a few packet losses or out-of-order deliveries should not pose a problem to the correctness of the final visual feedback. Since the primary function of the Irida system, in its current state, is to provide on-line feedback of the status and activity of a WSN testbed, we believe that no further effort is required on this front. Also, the UDP protocol is used only between the ICU and registered IVUs, which means that if more accurate ordering information is required in a particular scenario, it is always a possibility to develop synchronization algorithms alongside with the desired network functionality inside the WSN nodes firmware themselves.
\par
Many of the visualization commands that are available in the initial design of our system are scenario independent, such as the node heartbeat or packet exchange commands. Others can be used as the user see fit. For example, the color change command can be used to indicate clusters in a clustering algorithm, map temperature or radiation values to specific colors or any other similar use. Each commands requires at least one argument, which is the node ID. The node ID can be anything, from a number to the node's hex mac address, but it must be unique, since it identifies each node in the visualization. Uniqueness is not guaranteed by the system itself, so it must be designed into the hardware itself. We have used the node hex ID, which we manually verified to be unique in our specific testbed hardware.
\subsection{Visualization}
For proof of concept purposes, we have created an Irida Visualization Unit (IVU) which runs on Windows and uses XNA technology\cite{xna} for display. Since the Irida Control Unit (ICU) and IVU are decoupled and use a well established protocol with UDP sockets, IVU can be developed in any system and technology required to satisfy specific application needs. For example, the same technology can provide visualization for mobile devices (such as mobile phones or tablets) connected to the internet in another geographic locations. Our IVU runs full screen and the PC that runs it is connected to a projector. This projector then projects the WSN status in real time on top of the hardware nodes themselves. The IVU also provides some extra tools that are required for initial visual calibration such as rulers and grids. 
\par
Our IVU represents WSN nodes as coloured filled circles and the heartbeat animation as animated pulsing circles that grow and fade out when a heartbeat command is received. The position of the nodes is reported by the Irida enabled firmware on the nodes themselves. For messages exchange, our IVU display a small animated packet that moves white smaller circles from the message sender to the message receiver as well as a white line that shows the connection and fades out over time. WSN nodes also fade out over time, to denote inactive nodes when they haven't reported any activity (heartbeat or otherwise) over a configurable period of time. This provides visual feedback for nodes that were active before but seem to have switched to an inactive state (sleeping or waiting) at a later time. The neighbourhood links are also drawn as lines between nodes, that fade over a configurable period of time. This can be used to show links between nodes that have been confirmed as active in the past but haven't been used for a specified period of time. Of course, since nodes and links fading is configurable, this feature can be disabled if the "alive" status of these nodes and links is not important to the end user.
\par
Other information that a node wants to provide is also visualized by our IVU in the form of numerical badges and text labels. Each node draws a text node, typically with the node's name or ID, as well as two badges on the top left and top right side of the node, typically with numerical values. The badges content is application dependant, and can, for example, display message count or other information.
\par
Finally, our IVU can provide visual information about the WSN nodes neighbourhood. When nodes report adjacency information, in the form of neighbours, the IVU displays this information as lines that connect the nodes. In a similar fashion, when nodes report color information, the IVU adjusts their visual appearance to match this information. All the visual changes are performed using smooth fading animation but of course this is optional since developing an IVU only requires the implementation of visual feedback of the Irida Visualization Protocol commands.

\par

\section{Proof of concept algorithm}

\label{sec:algorithm}

For proof of concept reasons we have developed a simple algorithm that integrates the Irida protocol side to side with the algorithm functionality. The algorithm consists of two parts.

\par
The first part is a simple neighbourhood discovery algorithm. When a node turns on, as well as every 15+$a$ seconds, where $a$ is a random value in [0,5000] ms, it clears its known neighbours list, broadcasts a "Hello" message and waits for responses. Every node that receives such a hello message, compares its location with the location of the sender node, using a simple Manhattan distance measurement, and replies with a "Hello\_too" message if the distance is $\leq$ 2. In this case, the original sender node adds the responding node in its known neighbours list. This simple filtering allowed the network to produce small range neighbourhood information (as seen for example on fig.~\ref{fig:irida4x4_1}), as opposed to fully connected graphs which were possible due to the small size of the testbed.
\par
The second part consists of a sensor reactive notifier algorithm. When a sensor turns on, it records the sum of the 3 axis on the built-in accelerometer. Every time there is a change larger than a fixed threshold, a broadcast "forward" message is sent to all known neighbours. Each time a node receives a "forward" message, it queries a list of recent messages to see if it has seen this message before and if not, it forwards this message to its own list of neighbours. Using this simple mechanism, we have managed to flood this message in a controlled fashion while avoiding flooding loops at the same time.
\par
The firmware is based on a infinite loop consisting of three parts. At the beginning of the loop a node reads a entire packet and stores it in a list. The first part of the loop processes any stored packets and sends answers over the network. In the second part of the loop, the state of the node is verified and packets are sent if there are changes or if it is time to refresh the neighbourhood list. In the last part, the node listens for incoming packets. The node first waits for 1+$a$ seconds, where $a$ is a random value between [0,200] ms. If no data is received it restarts the loop. If it receives data, it reads it, stores it and waits for one additional second until there is no more data to read, then restarts the loop again. Packets can be lost while the system is in the first and second part of the loop because the firmware can not receive and send at the same time. Also, there's a minimum of one second between the reception of a packet and the sending of a reply.
\par  
Irida messages are sent each time something worth visualizing is happening. When any kind of message is received by any node, we send a "sendPacket" message to the ICU. Also, every few seconds we send a "heartBeat" message to signify that the node is active. Whenever a node is added in the known neighbours list, we send an "addNeighbour". When an accelerometer event is triggered or a node receives a "forward" message, we use "changeColor" to change the color of the node to red or blue accordingly.

\section{Conclusions and Future Work}
\label{sec:future}
In this paper, we described Irida, a real time visualization feedback protocol for Wireless Sensor Neworks. We argued the existence necessity of such a protocol for WSN algorithms research and demonstration and described the architecture choices we made when developing it. We have developed, deployed and tested our protocol on a physical WSN deployment to show its applicability.
\par
Even while developing a very simple algorithm like the one described in section~\ref{sec:algorithm}, the Irida protocol proved to be an extremely useful tool. Being an open and module based system, it was easy to integrate to our existing hardware platform and it helped fine tune and debug the algorithm in a point that it saved huge amounts of time and effort. Since the IVU modules are separate from ICUs, general purpose and ad-hoc implementations of IVUs are possible in the same system, which opens new possibilities for easy to understand demos of wireless sensor network applications.
\par
In future work, we plan to extend the system to take advantage of the existing data flow for more than just visualization. We intend to introduce a logging system that uses the Irida protocol packets to produce statistics and to allow storage of historical data for the activity of the network, for off-line study, statistics and evaluation of the running protocols.
\par
Our goal in the future is to make the ICU more active than just a flow controller. More specifically, we plan to introduce two major functions to the ICU:
\begin{itemize}
\item Off-line data storage and analysis. Since all of the status and activity messages pass through the ICU, it would be very useful to provide a mechanism to store them and analyse them off-line. For off-line analysis, it would become much more important to implement ways to ensure correct packets order in order to ensure in turn the correctness and re-playability of the executed scenarios.
\item Direct network control. In the current state of the Irida system, the ICU and IVU or IVUs are passive and the information flow always move from the network nodes to the ICU and from the ICU to the registered IVUs. We plan to introduce a parallel control protocol, that will enable the user to directly control some of the aspects of the network, like enabling and disabling sensors as well as execute commands (such as message routing or localization) with specific parameters and on demand. With the visualization feedback already in place, we feel that this would transform the Irida system into a more complete WSN testbed feedback and control system. 
\end{itemize}
Finally, we plan to extend the variety and functionality of our IVUs, to include more platforms (especially mobile).

\begin{appendices}
\section{Irida Visualization Protocol}
\label{sec:protocol}
The Irida protocol which is used both between the hardware and the ICU and between the ICU and IVU or IVUs is described in detail below:\par
\begin{tabular}{|p{7cm}|}
\hline
\textbf{heartBeat} \\
\hline
Required arguments:\\
\hline
$ID$ The unique ID of the node\\
Required arguments when used for the first time:\\
$x$,$y$ the position of the node in the network $\in$ [0,1]\\
\hline
example: heartBeat 0x00 0.5 0.5\\

This example introduces a new node with default settings at position 0.5, 0.5 if a node with id 0x00 does not already exist in the visualization or moves an existing node to the same position if it exists.\\
\hline
example: heartBeat $0x00$\\

This example initiates a heartBeat animation at the node with $0x00$ if it exists.\\
\hline
\end{tabular}

\begin{tabular}{|p{7cm}|}
\hline
\textbf{changeColor }\\
\hline
Required arguments:\\
\hline
$ID$: The unique ID of the node\\
$R$ $G$ $B$: the new color in R G B values $\in$ [0,255]\\
\hline
example: changeColor 0x00 255 0 0\\

This example changes the color of node with ID 0x00 to red.\\
\hline
\end{tabular}

\begin{tabular}{|p{7cm}|}
\hline
\textbf{activateNode }\\
\hline
Required arguments:\\
$ID$: The unique ID of the node\\
\hline
example: activateNode 0x00\\

This example activates the node with id 0x00 if it exists. Activated nodes are represented by default with no alpha value in the visualization\\
\hline
\end{tabular}

\begin{tabular}{|p{7cm}|}
\hline
\textbf{disactivateNode }\\
\hline
Required arguments:\\
$ID$: The unique ID of the node\\
\hline
example: disactivateNode 0x00\\
This example disactivates the node with id 0x00 if it exists. Disctivated nodes are represented by default with 50\% transparency in the visualization.\\
\hline
\end{tabular}

\begin{tabular}{|p{7cm}|}
\hline
\textbf{sendPacket }\\
\hline
Required arguments:\\
$IDsender$: The unique ID of the node which sends the packet\\
$IDreceiver$: The unique ID of the node which receives the packet\\
\hline
Optional arguments:\\
$Label$: An optional label to be attached to the data packet visualization\\

\hline
example: sendPacket 0x00 0x01\\
This example initiates a packet transfer animation between nodes with id 0x00 and id 0x01. \\
sendPacket 0x00 0x01 Data\\
This example initiates a packet transfer animation between nodes with id 0x00 and id 0x01. A "Data" label is added to the packet. \\
\hline
\end{tabular}

\begin{tabular}{|p{7cm}|}
\hline
\textbf{addNeighbor }\\
\hline
Required arguments:\\
$ID$: The unique ID of the node which a neighbour will be added to\\
$IDneighbor$: The unique ID of the neighbour node\\
\hline
example: addNeighbor 0x00 0x01\\
This example adds the node with id 0x01 to the node with id 0x01's neighbour list. Neighbourhood list is represented by lines drawn between neighbours in the visualization.\\
\hline
\end{tabular}

\begin{tabular}{|p{7cm}|}
\hline
\textbf{resetNeighbors }\\
\hline
Required arguments:\\
$ID$: The unique ID of the node which the neighbourhood list will be reset\\
\hline
example: resetNeighbors 0x00\\
This example empties the neighbourhood list of the node with id 0x00 if it exists.\\
\hline
\end{tabular}

\begin{tabular}{|p{7cm}|}
\hline
\textbf{setText }\\
\hline
Required arguments:\\
$ID$: The unique ID of the node\\
$Text$: The new text which will show in the visualization\\
\hline
example: setText 0x00 Cluster Leader\\
This example changes the text of 0x00 to "Cluster Leader".\\
\hline
\end{tabular}

\begin{tabular}{|p{7cm}|}
\hline
\textbf{setBadge }\\
\hline
Required arguments:\\
$ID$: The unique ID of the node\\
$BadgeNumber$: The ID of the badge to be changed (1 or 2)\\
$BadgeText$: The new text that the badge will display\\
\hline
example: setBadge 0x00 1 20\\
This example changes the badge number 1 text of 0x00 to "20".\\
\hline
\end{tabular}
\end{appendices}

{\small
\bibliographystyle{eg-alpha-doi}
\bibliography{tcssensewall}
}

\end{document}